\documentclass{iopart} 
\usepackage{xspace}
\usepackage{url}

\usepackage{graphicx}

\usepackage{color}

\let \IG \includegraphics

\begin{document}

\title[Coupling of lateral grating displacement] 
{Coupling of lateral grating displacement to the output ports of a 
 diffractive Fabry-Perot cavity}

\author{J.~Hallam$^1$, S.~Chelkowski$^1$, A.~Freise$^1$, S.~Hild$^1$, B.~Barr$^2$, K.A.~Strain$^2$, O.~Burmeister$^3$
and R.~Schnabel$^3$} 
\address{$^1$ School of Physics and
Astronomy, University of Birmingham, Edgbaston, Birmingham B15 2TT, United Kingdom}
\address{$^2$ Institute for Gravitational Research, Department of Physics and Astronomy, University of Glasgow, Glasgow, G12 8QQ, United Kingdom} 
\address{$^3$ Max-Planck-Institut f\"ur Gravitationsphysik
(Albert-Einstein-Institut) and Leibniz Universit\"at Hannover,
Callinstr.~38, D--30167 Hannover, Germany.}

\ead{jmh@star.sr.bham.ac.uk}

\begin{abstract}

Diffraction gratings have been proposed as core elements
in future laser-interferometric gravitational-wave detectors. In this paper, we use
a steady-state technique to derive coupling of lateral  grating displacement 
 to the output ports of a
 diffractive Fabry-Perot cavity. By introducing a signal to noise ratio (SNR) for each of the three cavity
 output ports the magnitude of the noise sidebands originating from lateral
grating displacement are
  compared to the magnitude of a potential gravitational wave signal. 
For the example of a 3km long Fabry-Perot cavity featuring parameters similar 
to the planned Advanced Virgo instrument, we found that the forward-reflecting
grating port offers the highest SNR at low frequencies. Furthermore, for
this example suspension  requirements for lateral isolation were computed, and a 
factor of twenty relaxation at a frequency of 10\,Hz can be gained over the transmitted
port by observing the forward-reflected port. 

\end{abstract}


\submitto{\JOA}
\maketitle


\section{Introduction}
\label{sec:intro}

It has been suggested that the sensitivity of high-precision
 interferometers, with particular reference to laser-interferometric 
 gravitational wave detectors, could be improved by using an all-reflective 
optical configuration \cite{Sun98}. Expected improvements include eliminating transmissive thermal
 noise and freeing optical substrates to be opaque thus widening the
 range of material choices. Furthermore, the shot noise contribution can be
 reduced by increasing laser power without increasing technical problems such as thermal lensing.
 Diffractive optics have been suggested as possible replacements for mirrors 
and beam-splitters that could function in an all-reflective configuration \cite{Drever96}. This document
 considers the case where a diffractive optic is used in place of a mirror to couple light 
into a Fabry-Perot cavity \cite{Bunkowski04}. 

It has been found that even state of the art diffraction gratings are
 too low in diffraction efficiency to allow high-finesse cavities based on the grating diffraction \cite{Bunkowski06c, Perry95}.
 An alternative is to use a diffraction grating with low diffraction efficiency as an incoupler to a
 Fabry-Perot cavity, because in this case there is the advantage 
that the cavity finesse is determined by the reflection
 efficiency of the grating. 
In three-port, second-order-Littrow configuration (as shown in Figure \ref{fig:3p_grating_movec})
 it is the zeroth-order diffraction (equivalent to the reflection)
 that retains the light-field in the Fabry-Perot cavity \cite{Bunkowski04}.
 Standard dielectric coating techniques 
 provide the high-reflectance efficiency required for such a configuration.

\section{Phase noise originating from lateral oscillation of grating}
\label{sec:origins}

In comparison to a standard Fabry-Perot cavity featuring a transmissive 
input mirror the diffractive nature of a grating causes an additional coupling
 of geometry changes into phase noise \cite{Freise07}. This phase noise 
stems purely from the geometry of the optics, and can be fully calculated 
from the grating equation and the geometry of the set-up alone. Freise et al 
determined that, assuming a sensitivity goal of $h = 10^{-23}/\sqrt{{\rm Hz}}$,
 Virgo-like interferometer parameters \cite{acernese08} and typical values for
 the grating parameters, a standard cavity required end mirror angular alignment
 of $\gamma < 2\cdot10^{-16}{\rm \,rad}/\sqrt{{\rm Hz}}$ whereas a three-port
 coupled grating cavity required $\gamma < 1\cdot10^{-21}{\rm \,rad}/\sqrt{{\rm Hz}}$,
 needing a five orders of magnitude improvement in the end turning mirror 
alignment for equivalent insensitivity to this noise source. 
Improvement in alignment of the injection optics and lateral grating suspension would also be required.

\begin{figure}[Htb]
\begin{center}
\IG [scale=0.25]{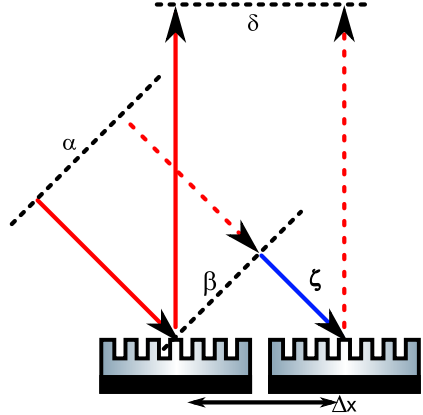}
\end{center}
\caption{Wavefronts coupling through a diffraction grating.
 The wavefront (propagation vector shown in solid red) originates at $\alpha$, 
travels to $\beta$ and is diffracted to $\delta$. When the grating is laterally 
displaced by $\Delta x$, the wavefront (propagation vector shown
 red-dashed) must pass through additional path length $\zeta$ (propagation 
vector shown in solid blue).}
\label{fig:pc}
\end{figure}

In this article we consider the similar case of grating displacement perpendicular
 to the beam and across the grating corrugation (lateral displacement) which couples into phase noise. The phase 
change resulting from the displacement is due to a path-length difference $\zeta$ between parallel 
wavefronts, shown for a simple geometrical case in Figure \ref{fig:pc}.


\section{Frequency domain modelling of phase noise coupling in a grating cavity}
\label{sec:intro2}

In order to quantitatively analyse the coupling of the phase noise introduced by grating displacement to the output ports
of a diffractive cavity we perform a frequency domain
 analysis. By carrying out this analysis for the effect of a potential gravitational-wave
 signal and the noise effect of lateral grating 
 displacement we can obtain the signal to noise ratio (SNR) at all three output 
ports of the grating cavity. The aim is to find the
 port with the best SNR.

To compare the signal to noise ratio at the different output ports we first determine the grating coupling relations.
We then review the mechanism of frequency sideband generation by a laterally
 displaced grating and identify the complex field amplitudes at the sideband frequency caused by 
 (and in terms of) each carrier light field incident
  on the grating. 
  Using the coupling relations we calculate each carrier light field
   incident on the grating in terms of a single external input carrier field. These are then used to determine the amplitude of the input sideband
    fields. Again utilising the coupling relations we obtain the amplitude of the sideband fields at the output ports.
     We next derive the amplitude of the potential gravitational wave signal sidebands at the output ports and 
     divide by the grating displacement noise sidebands to obtain the signal to noise ratio.
     
The typical expression for an electromagnetic light field is $E = a_q e^{-i \omega t}$, 
where $a_q$ is the field amplitude (which may be a complex number) with
$q$ a numeric subscript denoting location in the optical system consistent with Figure \ref{fig:3p_grating_movec} 
and $\omega$ is the angular frequency of the light field. 
For coupling relations we will use $a_q$ instead of $E$. The
frequency $\omega$ of a field described by an amplitude $a_q$ is
determined by the context.


\subsection{Coupling relations of a static grating}
\label{sec:FSG}

\begin{figure}[Htb]
\begin{center}
\IG [scale=0.25]{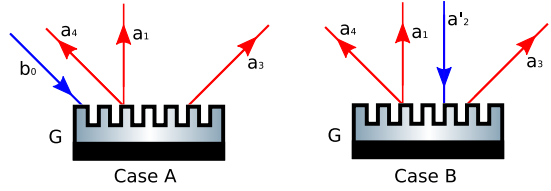}
\end{center}
\caption{Input ($b_0$, $a_2'$, shown blue) and output ($a_1$, $a_3$, $a_4$, shown red) light field amplitudes at a phase modulating grating in second-order Littrow-configuration in case A, and zeroth-order-Littrow configuration in case B (labels consistent with Figure \ref{fig:3p_grating_movec}).}

\label{fig:FA}
\end{figure}




The coupling relations of a single, static grating are given in \cite{Bunkowski05}:

\begin{equation}
\label{nuG3Pi}
\left( \begin{array}{ccc}
a_4 \\
a_1 \\
a_3 \end{array} \right)
 =  \left[ \begin{array}{ccc}
\eta_2 e^{-i\phi_2} & \eta_1 e^{-i\phi_1} & \eta_0 \\
\eta_1 e^{-i\phi_1} & \rho_0 & \eta_{1} e^{-i\phi_{1}} \\
\eta_0 & \eta_{1} e^{-i\phi_{1}} & \eta_{2} e^{-i\phi_{2}} \end{array} \right]
\left( \begin{array}{ccc}
b_0  \\
a_2'  \\
0 \end{array} \right).
\end{equation}

\noindent Where $\eta$ is the amplitude coefficient and $\phi$ is the constant 
phase change associated with diffraction in the order given by the subscript. Bunkowski et al \cite{Bunkowski05}
have shown that positive and negative diffraction orders have the same $\eta$ and $\phi$ in this configuration
so only positive subscripts will be used.
Zeroth order diffraction is reflection and $\rho_0$ is the amplitude
 coefficient in the special case of zeroth order diffraction normal to the surface 
of the grating. Adopting simplified notation we can write:

\begin{equation}
\label{G3P}
\left(\frac{}{} \begin{array}{ccc}
a_4 \\
a_1 \\
a_3 \end{array} \right)
 =  \left[ \begin{array}{ccc}
G_2 &G_1& G_0\\
G_1 & \rho_0 & G_{1} \\
G_0 & G_{1} &G_{2} \end{array} \right]\left(\frac{}{} \begin{array}{ccc}
b_0  \\
a_2'  \\
0 \end{array} \right).
\end{equation}

\subsection{Coupling relations of an oscillating grating}
\label{sec:CROG}

Consider the case of an oscillating grating with two incident light fields shown in Figure \ref{fig:FA}:
In case A the external input field with amplitude $b_0$ 
and in case B the field with amplitude $a_2'$.
The frequency sideband generation for a grating oscillating with 
angular frequency $\omega_m$ and modulation index $m$ will
be demonstrated for output $a_1$ in case A. The oscillating grating causes
phase modulation of the diffracted fields \cite{Freise07}:

\begin{equation}
a_1 =  b_0 G_1 e^{-i m \cos (\omega_m t)}.
\end{equation}

\noindent The complex field amplitude resulting can be expanded using the Bessel function $J_k(m)$ \cite{finesse0998, phd.Freise}. In the 
case $m \ll 1$ it is sufficient to consider terms linear in $m$. 
It follows that only $k = -1,0,1$ need to be considered resulting in coefficients:

\begin{eqnarray}
J_1(m) = m/2,\qquad J_0(m) = 1,\qquad J_{-1}(m) = -m/2 \label{besel1}.
\end{eqnarray}

\noindent Thus obtaining:

\begin{equation}
a_1 =  b_0 G_1 \left(J_0(m) + i J_{1}(m) e^{i \omega_m t}  - i J_{-1}(m) e^{-i \omega_m t}\right).
\end{equation}
\begin{equation}
a_1 =  b_0 G_1 \left(1+ i \frac{m}{2} e^{i \omega_m t}  - i \frac{m}{2} e^{-i \omega_m t}\right).
\end{equation}

\noindent When a grating is laterally oscillated to a maximum displacement of $\Delta x$ the modulation 
index is given by $m = 2 \pi \Delta x \mu/d$, where $d$ is the corrugation period of the 
grating and $\mu$ is the diffraction order, resulting in \cite{Freise07}:

\begin{equation}
\label{seven}
a_1 =  b_0 G_1 \left(1+ i \frac{ \pi \Delta x \mu}{d} e^{i \omega_m t}  - i \frac{ \pi \Delta x \mu}{d} e^{-i \omega_m t}\right).
\end{equation}

\noindent The terms including $e^{\pm i \omega_m t}$ are called sidebands. In the following we will compute the sideband amplitudes in a cavity with the grating as the input coupler. The amplitudes in front of the $e^{\pm i \omega_m t}$ as computed in Equation \ref{seven}, when coupled through the grating, are the sideband amplitudes for output $a_1$ in case A \cite{phd.Heinzel}, \cite{Mizuno99}. First order diffraction ($\mu = 1$) from the grating is the coupling between $b_0$ and $a_1$:

\begin{eqnarray}
\label{duo}
i b_0 \frac{\pi \Delta x}{d} G_1 \label{ba}.
\end{eqnarray}

\noindent Having demonstrated the calculation of the sideband amplitude for output $a_1$ in case A we will now write the result for outputs $a_3$ and $a_4$ in case A using the same method. Zeroth order diffraction ($\mu = 0$) occurs between $b_0$ and $a_3$ and therefore no sidebands are generated. Second order diffraction ($\mu = 2$) occurs between $b_0$ and $a_4$ so the sideband amplitude obtained is:

\begin{eqnarray}
\label{duo3}
i b_0 \frac{2 \pi \Delta x}{d} G_2 \label{bz}.
\end{eqnarray}


\noindent Having dealt with case A we proceed to case B in Figure \ref{fig:FA}. Zeroth order diffraction ($\mu = 0$) occurs between $a_2'$ and $a_1$ and therefore no sidebands are generated. Negative-first order diffraction ($\mu = -1$) occurs between $a_2'$ and $a_3$ so the sideband amplitude obtained is: 

\begin{eqnarray}
\label{trio}
- i a_2' \frac{ \pi \Delta x}{d} G_1 \label{aa}.
\end{eqnarray}

\noindent First order diffraction ($\mu = 1$) occurs between $a_2'$ and $a_4$ so the sideband amplitude obtained is:

\begin{eqnarray}
\label{triob}
i a_2' \frac{ \pi \Delta x}{d} G_1\label{az}.
\end{eqnarray}



\subsection{Coupling relations of an oscillating grating as Fabry-Perot cavity input optic}
\label{sec:couprel}

\begin{figure}[Htb]
\begin{center}
\IG [scale=0.25]{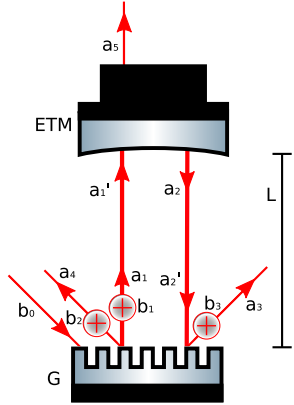}
\end{center}
\caption{External input port $b_0$, internal input ports $b_1$, $b_2$, $b_3$ (marked by encircled addition symbols) and transfer ports $a_{q}$ in a grating cavity.}
\label{fig:3p_grating_movec}
\end{figure}

To determine the transfer function for the generated noise sidebands to the output ports $a_3$, $a_4$ and $a_5$, and the input carrier 
field to ports $b_1$, $b_2$ and $b_3$ (in Figure \ref{fig:3p_grating_movec}) it is necessary to know the port to port coupling relations 
for our three-port coupled diffractive cavity of interest. Hence
the coupling relations for the end mirror ETM with amplitude transmission coefficient $\tau$ and amplitude 
reflection coefficient $\rho$ must be known:

\begin{equation}
\label{ETM}
\left(\frac{}{} \begin{array}{cc}
a_2 \\
a_5 \end{array} \right)
 =  \left[ \begin{array}{cc}
\rho_2 &i \tau_2\\
i \tau_2 & \rho_2\end{array} \right]\left(\frac{}{} \begin{array}{cc}
a'_1  \\
0 \end{array} \right),
\end{equation}

\noindent and for the space of length $L$ between the grating and mirror:

\begin{equation}
\label{length}
\left(\frac{}{} \begin{array}{cc}
a_1' \\
a_2' \end{array} \right)
 =  \left[ \begin{array}{cc}
e^{-ikL} & 0 \\
0 & e^{-ikL}\end{array} \right] 
\left(\frac{}{} \begin{array}{cc}
a_1  \\
a_2 \end{array} \right),
\end{equation}

\noindent where $k = \omega/{\rm c}$, with $\omega$ the angular frequency 
and c the speed of light. Hence it is possible to write the local coupling relations
for the diffractive cavity shown in Figure \ref{fig:3p_grating_movec} including the internal inputs discussed in Section \ref{sec:CROG} at ports
$b_1$, $b_2$ and $b_3$:

\begin{eqnarray}
a_{{\rm1}} &=& b_{0} G_1 + a'_{{\rm2}}  \rho_0 \, + b_{{\rm1}} \label{nui} \nonumber \\
a'_{{\rm1}} &=& a_{{\rm1}} e^{-ikL}    \label{nuii} \nonumber \\
a_{{\rm2}} &=& a'_{{\rm1}} \rho_2 \label{nuiii} \nonumber \\
a'_{{\rm2}} &=& a_{{\rm2}} e^{-ikL} \label{nuiv} \nonumber\\
a_{{\rm3}} &=& b_{0} G_0 + a'_{{\rm2}} G_{1} \, + b_{{\rm3}}  \nonumber\label{nuv}\\
a_{{\rm4}} &=& b_{0} G_2 + a'_{{\rm2}} G_1 \, + b_{{\rm2}}  \label{nuvi} \nonumber\\
a_{{\rm5}} &=& a'_{{\rm1}} i \tau_2  \label{nuvii}.
\end{eqnarray}


\noindent For the purposes of simplification, we define the resonance term of the cavity:

\begin{equation}
D(\omega) = \frac{1}{1 - \rho_2 \rho_0 e^{-2ikL}}\label{nuzaaii}.
\end{equation}

\noindent  By performing some substitution we obtain the port to port relations only in terms of the input ports $b$ and the fixed cavity parameters:

\begin{eqnarray}
a_{{\rm1}} &=&  (b_{0} G_1 +  b_{{\rm1}}) D(\omega) \label{nuxi}  \nonumber\\
a'_{{\rm1}} &=& (b_{0} G_1 +  b_{{\rm1}}) D(\omega)  e^{-ikL} \label{nuxii} \nonumber \\
a_{{\rm2}} &=& (b_{0} G_1 +  b_{{\rm1}}) D(\omega)  \rho_2 e^{-ikL}  \label{nuxiii}  \nonumber\\
a'_{{\rm2}} &=& (b_{0} G_1 +  b_{{\rm1}}) D(\omega)  \rho_2 e^{-2ikL} \label{nuzxiv} \\
a_{{\rm3}} &=& b_{0} G_0 + (b_{0} G_1 +  b_{{\rm1}}) D(\omega)  \rho_2 e^{-2ikL} G_{1} \, + b_{{\rm3}} \label{nuzxv}\\
a_{{\rm4}} &=& b_{0} G_2 + (b_{0} G_1 +  b_{{\rm1}}) D(\omega)  \rho_2 e^{-2ikL} G_1 \, + b_{{\rm2}}  \label{nuxvi}\\
a_{{\rm5}} &=& (b_{0} G_1 +  b_{{\rm1}}) D(\omega)  e^{-ikL} i \tau_2  \label{nuzxvii}.
\end{eqnarray}


\subsection{Carrier field solution}
\label{sec:cfs}

\noindent From the carrier amplitude present at $a_2'$ we can determine the sideband amplitudes at $b_1$, $b_2$ and $b_3$. To obtain the carrier amplitude at $a_2'$ it is necessary to solve Equation \ref{nuzxiv} with an example input field at port $b_0$ of arbitrary angular frequency and amplitude $p_0 e^{-i \omega_c t}$ and no field at the internal input ports $b_1$, $b_2$, $b_3$, hence obtaining:

\begin{eqnarray}
a'_{2} &=&  p_0 G_1 D(\omega_{c}) \rho_2 e^{-2ik_cL}  \label{nuzxivb},
\end{eqnarray}

\noindent where $a'_2$ is the complex field amplitude of the carrier field. 
Since this solution is specific to the carrier, we distinguish it by using the subscript $c$, and perform some simplification by introducing $B_c =  D(\omega_c) \rho_2 e^{-2ik_cL}$:

\begin{eqnarray}
a'_{2} &=&  p_0 G_1 B_c \label{nuzzlexiv}.
\end{eqnarray}

\subsection{Input sideband fields}

There is now sufficient information to determine the sideband field amplitudes 
$b_1$, $b_2$ and $b_3$. These are the linear sums of the sidebands created by the oscillating
grating with the impinging fields in $b_0$ and $a_2'$. Using Equations \ref{ba} through \ref{az} and \ref{nuzzlexiv}
we can now write:

\begin{eqnarray}
b_{1} &=& i p_0 G_1 \frac{\pi \Delta x}{d}\label{b1}, \\
b_{2} &=& i p_0 G_2 \frac{2 \pi \Delta x}{d} + i p_0 G_1^2 B_c \frac{\pi \Delta x}{d} \label{b2}, \\
b_{3}&=& - i p_0 G_1^2 B_c \frac{\pi \Delta x}{d} \label{b3},
\end{eqnarray}

\noindent where an input field of the amplitude $p_0$ at port $b_0$ has been used.



\subsection{Sideband fields at outputs}

To determine the sideband field amplitude at the outputs $a_3$, $a_4$ and $a_5$ we consider the total field at the outputs (given by Equations \ref{nuzxv} through \ref{nuzxvii}). Setting the external input at port $b_0$ to zero leaves the internal inputs which are expressions of the sidebands. Equations \ref{b1} through \ref{b3} for the internal inputs are substituted into these output field equations allowing them to be compared in terms of the fixed cavity properties and $p_0$. The sideband fields have different frequencies ($\omega_u = \omega_c + \omega_m$ for 
the upper sideband, $\omega_l = \omega_c - \omega_m$ for the lower sideband) and hence we address them separately, upper sideband first:

\begin{eqnarray}
a_{{\rm3}} &=& i p_0 G_1^2 \frac{\pi \Delta x}{d} \left(B_{u} - B_c \right)\label{nubxv},\,\\
a_{{\rm4}} &=& i p_0 G_1^2 \frac{\pi \Delta x}{d} \left(B_{u} + B_c \right) + i p_0 G_2 \frac{2 \pi \Delta x}{d}, \label{nubxvi} \\
a_{{\rm5}} &=& - p_0 G_1 \frac{\pi \Delta x}{d} \frac{B_{u}}{\rho_2 e^{-ik_{u}L}} \tau_2 . \label{nubxvii}
\end{eqnarray}

\noindent $a_3$, $a_4$ and $a_5$ are complex amplitudes associated with the upper sideband at each output, whilst $p_0$ is the complex amplitude for the carrier field input in the port $b_0$. $\omega_{u}$ is the absolute frequency of the upper sideband and 
$B_{u} =  D(\omega_{u})\rho_2 e^{-ik_{u}L}$, analogous to our previously defined $B_c$. The calculation can be repeated to show 
that the lower sideband has the same magnitude as the upper at all output ports. We have thus computed the optical signal in all output ports generated by lateral motion of the grating.

\section{Signal coupling in a grating cavity} 
\label{sec:gw}


\noindent In this section we consider the interaction between the cavity and a potential gravitational wave
 propagating perpendicular to the cavity axis. The gravitational wave
 imposes phase modulation sidebands onto the light-field inside the cavity. 
 
 The modulation index of the gravitational wave can be expressed in terms of an equivalent
 displacement of the mirror $\Delta z$ resulting in modulation index $m = 4 \pi \Delta z/\lambda$.
 The effect of this displacement appears at cavity internal
 input $b_1$ in Figure \ref{fig:3p_grating_movec} with no input in ports $b_2$ and $b_3$. 
 The port-to-port coupling relations given in Section \ref{sec:couprel}, and specifically
 Equations \ref{nuzxv} through \ref{nuzxvii} for the output ports, can be used since the cavity coupling relations are 
 unaffected by the case chosen (displaced grating or incident gravitational wave).
 
 The carrier field amplitude $a_2'$ can still be used from Equation \ref{nuzzlexiv} as the carrier field is by definition unaffected when changing from the displaced grating to incident gravitational wave case. Hence from the
 modulation index and the carrier field present, again using the example input field of amplitude $p_0$, we obtain the internal input
 amplitude:
 
 \begin{eqnarray}
 b_1 = i p_0 G_1 B_c \frac{ 2 \pi \Delta z}{\lambda}.
 \end{eqnarray}
 
 \noindent Therefore the output amplitudes at the upper sideband frequency are:
 
 \begin{eqnarray}
 a_{{\rm3}} &=& i p_0 G_1^2 \frac{2 \pi \Delta z}{\lambda} B_c B_u \label{ncbxv},\\
 a_{{\rm4}} &=& i p_0 G_1^2 \frac{2 \pi \Delta z}{\lambda} B_c B_u \label{nucxvi}, \\
 a_{{\rm5}} &=& - p_0 G_1 \frac{2 \pi \Delta z}{\lambda} \frac{B_c}{\rho_2 e^{-ik_cL}} \frac{B_u}{\rho_2 e^{-ik_uL}} \tau_2 \label{nucxvii}.
 \end{eqnarray}
 
 \noindent We have thus computed the optical signal in all output ports due to a potential gravitational wave signal.
 
\section{Ratio of signal to noise at the outputs of a grating cavity}
\label{Sec:SNR}


The ratio of gravitational wave signal to lateral grating displacement noise (the signal to noise ratio) 
will be used as a figure of merit to evaluate the interferometric length sensing 
performance of the different diffractive cavity output ports. 
In order to derive the SNR the  absolute field amplitude 
in the gravitational wave case will be divided by the absolute field amplitude in the grating lateral 
displacement case for each output\footnote{Note that for a standard cavity
 with equivalent finesse, the SNR will be infinity as in principle
 two-mirror cavity input mirrors are insensitive to lateral displacement. In reality however in any suspension system there is always a coupling from lateral excitation to longitudinal displacement.}. 
Using Equations \ref{nubxv} through \ref{nubxvii} and \ref{ncbxv} 
through \ref{nucxvii} we obtain the following signal to noise ratios at the output port of the equivalent subscript:



\begin{eqnarray}
{\rm SNR_3}  &=&\Lambda_{\rm cav} \frac{ B_{u}}{B_{u} - B_{c}}, \label{xxti} \\
{\rm SNR_4}  &=&\Lambda_{\rm cav} \frac{B_{u}}{B_{u} \, +  B_{c} + 2 G_2/G_1^2}
 \label{xxtii} , \\
{\rm SNR_5} &=&\Lambda_{\rm cav}/\rho_2 e^{-ik_{c}L} \label{xxtiii},
\end{eqnarray}

\noindent with

\begin{equation}
\Lambda_{\rm cav} = \frac{2 d \Delta z B_{c}}{\lambda \Delta x}.  \label{caplamb} 
\end{equation} 

\noindent These equations reveal the differences between the  ratio of 
gravitational wave signal to lateral grating displacement noise for the three 
cavity output ports. This SNR is frequency independent for the case of the output port in transmission of the
cavity end mirror ($a_5$), while the SNR at the other two output ports contain the
modulation frequency dependent term $B_{u}$.


It was found that the SNR for the two output ports $a_3$ and $a_4$ includes the SNR
from inside the cavity $\Lambda_{\rm cav}$, multiplied by a fraction containing
 the different resonance
factors for the carrier light field and the sidebands. Apart from the small $G_2/G_1^2$ term the only 
 difference between Equations
\ref{xxti} and \ref{xxtii} is whether the resonance terms of the carrier and sideband in 
the denominator are added or subtracted. The important feature of these equations
is that $B_u \approx B_{c}$ for small modulation frequencies (i.e. modulation frequencies within the cavity
bandwidth). This brings the denominator in
Equation \ref{xxti} close to zero, thus strongly increasing the SNR
 at port $a_3$, i.e. resulting in
a partial cancellation of the phase noise introduced from lateral grating displacement.

\subsection{Numerical result}
\label{Sec:SNRNR}

In this section we present a quantitative 
analysis for one 
example configuration.  We choose a cavity length of  $L = 3\,{\rm km}$,
$\rho_2 = \sqrt{0.99995}$ (50\,ppm power transmittance) and  $\rho_0 = \sqrt{0.95}$.
Demanding that the $\mu = 1$ diffraction order propagates normally to the grating for the input
 beam imposes the grating design requirement $d \leq 2 \lambda$ \cite{Clausnitzer05}.
 The grating phase relations further impose the minimum possible value of 
$\eta_2 \geq 0.0127$.  From \cite{Bunkowski05} 
for minimum $\eta_2$ we find that $G_2 = -0.0127$. 
$G_1$ is fixed by the required cavity input parameters.
Together $G_1$ and $G_2$ set the value of the last term in the denominator in Equation \ref{xxtii} 
and it is small compared to the other terms of the denominator. 
The cavity is set to resonance for the carrier light, imposing $e^{-2ik_{c}L} = 1$ and the 
modulation frequency $\omega_m$ of the lateral grating displacement is chosen to be 10\,Hz.

Our interest focuses on the relative magnitude of the frequency sidebands. As in Freise et al \cite{Freise07} 
we set $\Delta z/\Delta x = 1$ in order to compare potential gravitational-wave signal to the effect of 
grating displacement.
 Since we are dealing with complex field amplitudes for this analysis we take the
 absolute value of the ratios\footnote{The ratio contains a real part as well
as an imaginary part. We assumed that the readout quadrature of each of 
the three output ports can be chosen individually by adding a proper
local oscillator, i.e. performing a homodyne measurement.} 
given in Equations \ref{xxti} to \ref{xxtiii}, yielding the following result:

\begin{eqnarray}
 \left|{\rm SNR_3}\right| &=& 3181,\\
 \left|{\rm SNR_4}\right| &=& 79,\\
 \left|{\rm SNR_5}\right| &=& 158. \label{a5num}
\end{eqnarray}

In all output ports the SNR is found to be greater than one, 
as the cavity suppresses the displaced grating phase 
noise sidebands with a factor of 158 as seen in the transmitted output port; 
however, a factor of twenty improvement in the SNR can be obtained through the cancellation of 
input and output grating displacement sidebands in the forward-reflected ($a_3$) port compared to the transmitted ($a_5$) port.

For proposed interferometric gravitational wave detector layouts the end-mirror will be highly reflective ($\rho_2$ will be close to one) and therefore the complex field amplitudes for the potential gravitational wave signal (Equations \ref{ncbxv} through \ref{nucxvii}) will obey the relations $a_3 \gg a_5$, $a_4 \gg a_5$. There will be approximately half the signal in the forward-reflected port with good SNR, half will be in the back-reflected port with poor SNR and 
very little will be in the transmitted port. 

\begin{figure}[Htb]
\centering
\IG [scale=0.75] {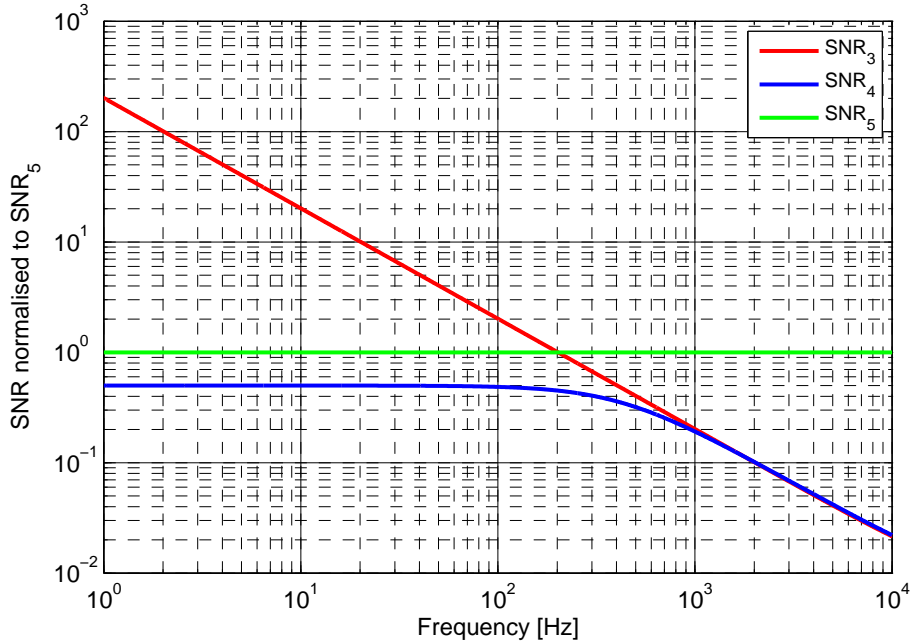}
\caption{Signal to noise ratio of gravitational wave equivalent mirror displacement
and lateral grating displacement at the three different output ports of a  grating gravity. All
traces are normalized to ${\rm SNR_5}$.} 
\label{fig:GratSNR}
\end{figure}

In a next step, instead of a single frequency of interest, we consider modulation
frequencies covering the full detection band of gravitational wave detectors.
 As ${\rm SNR_3}$ and ${\rm SNR_4}$ are
 frequency dependent, we normalise them to ${\rm SNR_5}$ and 
plot them over frequency to obtain Figure \ref{fig:GratSNR}.  

At low modulation frequencies where the sidebands are to a good approximation 
resonant in the cavity, imposing $B_{u} \approx B_{c}$ the forward-reflected port ($a_3$) has
 good cancellation between the input and output sidebands generated by the grating
 lateral displacement. This results in 
high SNR compared to the transmitted port ($a_5$). In the back-reflected port ($a_4$) summation occurs instead of cancellation
  and hence the SNR is lower than in the transmitted port. As the
 modulation frequency diverges from the cavity resonance, the sidebands
 do not resonate in both the gravitational-wave and grating lateral displacement 
case as $B_{u}$ trends to zero. Thus the sideband contribution generated when the carrier field exits the cavity, the $B_{c}$ terms in the denominator of Equations \ref{xxti} and \ref{xxtii},
 dominates causing the SNR of the reflected ports to converge below the level
 of the transmitted port.

\subsection{Suspension requirements}

\begin{figure}[Htb]
\centering
\IG [scale=0.75] {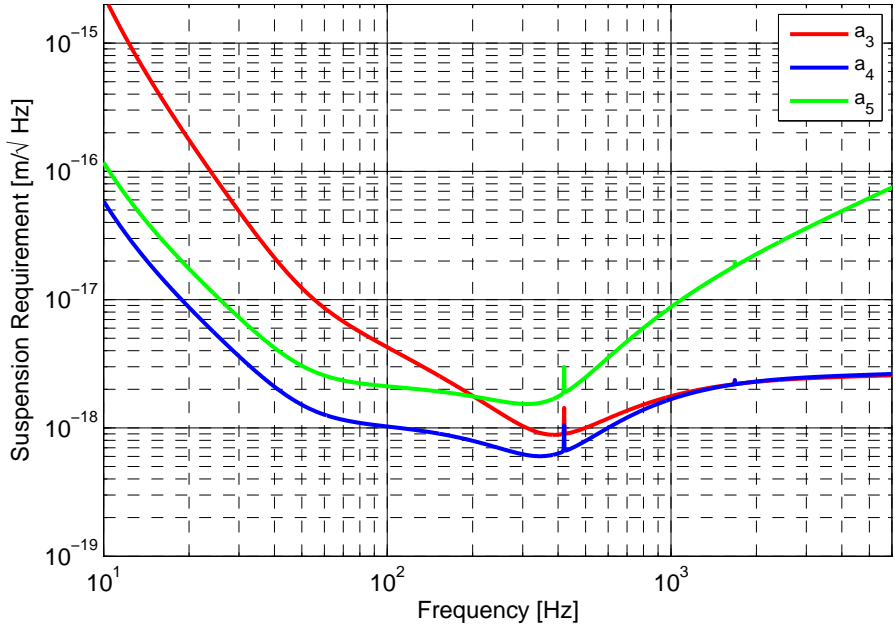}
\caption{Suspension requirement for the maximum tolerable lateral grating displacement that is necessary to
achieve the Advanced Virgo design sensitivity. The suspension required strongly depends
 on the actual readout port as well as the frequency of interest. At low frequencies the
 phase noise suppression in port $a_3$ allows significant relaxation of the
required suspension isolation. The narrow peak around 410\,Hz is due to a resonance in the Advanced Virgo design,
not a grating effect.}
\label{fig:SusReq}
\end{figure}

\noindent To ease comparison of the results derived above we determine 
the potential suspension requirements for a grating cavity used
as an arm cavity within the planned Advanced Virgo detector.
Using the current Advanced Virgo design sensitivity \cite{VIR-101A-08}
we plot in Figure \ref{fig:SusReq} the corresponding tolerable lateral grating motion for each
of the three potential readout ports of the grating arm cavity.



 We have found that we can relax by a factor of twenty at 10\,Hz (Figure \ref{fig:SusReq}) the suspension requirement for the the 
 un-suppressed transmitted ($a_5$) port by utilizing the $a_3$ (forward-reflected) port for the signal readout. 
 This suppression depends on cancellation between noise sidebands which we have calculated for the case of
 the lateral suspension of the grating. It is likely that this suppression will also apply to the input pointing (as misaligned input pointing also
 changes the alignment of the field circulating in the cavity \cite{Morrison1994b}) and hence the alignment stability requirement for the injection optics.
 However sidebands due to end-mirror angular misalignment are not suppressed.
 
 Freise et al \cite{Freise07} found that without suppression of the grating lateral displacement using a three-port-coupled
 grating cavity rather than a standard two-mirror cavity, in the Advanced Virgo case, requires improving the end-mirror angular alignment
 suspension requirement by five orders of magnitude. The equivalent improvement required for the injection optics 
 alignment stability and grating lateral suspension can be relaxed by a factor of twenty by choosing the forward-reflected 
 port for the signal readout, at least in the lower frequency band.
 
 

 Above 200\,Hz this geometrical effect no longer suppresses
 coupling of lateral grating displacement in the forward-reflected port. 
 In this case it is preferred to use the transmitted port for readout; however,
 the fixed SNR set by the cavity properties is not beaten, and the system is no longer entirely reflective.

 
 The  encouraging results presented above increase
 the application prospects of grating coupled Fabry-Perot cavities
 in large-scale gravitational wave detectors, but the problem 
 raised by Freise et al of end-mirror angular alignment remains.

\section{Summary and Outlook}

Diffractive optics are an important avenue of research because 
they allow all-reflective interferometer configurations which offer 
thermal noise improvements and may be especially relevant in
 mitigating effects made important when high power is used to 
reduce shot-noise contribution. An additional noise source is 
 lateral grating displacement relative to the beam causing path-length differences 
that couple phase-noise into the gravitational wave detection channel.
 From \cite{Freise07} the grating-cavity end-mirror angular alignment suspension 
requirement (five orders of magnitude higher than that required 
by a two-mirror cavity), and associated likely increases in injection optic pointing and lateral grating stability 
were a significant impetus against using grating cavities. 

 The diffraction order ($\mu$) dependence of the frequency sideband amplitude
 and the symmetry inherent in the three-port-coupled grating cavity 
 suggested that the phase noise generated on input to the cavity might
 cancel with the phase noise generated at output from the cavity.
 
 To determine the effect we carried out a frequency 
domain analysis to calculate the coupling of lateral grating displacement
to the different output ports of a
 three-port-coupled grating cavity. 
For the output port in forward-reflection of the grating we found
a suppression of phase noise originating
from lateral grating displacement over the transmitted port, resulting in a factor of twenty relaxation 
in the lateral displacement suspension requirement at 10\,Hz. This will likely also apply to the injection optics pointing stability. 
This factor will increase with a wide cavity bandwidth and hence the noise suppression will be greater in cavities of lower finesse.


\section{Acknowledgements}

This work was supported by the Science and Technology Facilities Council (STFC), the European Gravitational Observatory (EGO) and the Deutsche Forschungsgemeinschaft (DFG) within the Sonderforschungsbereich (SFB) TR7.

\section*{References} 



\bibliographystyle{unsrt}
\bibliography{bham-grating}

\end{document}